\documentclass[aps,superscriptaddress,nofootinbib,twocolumn]{revtex4}
\usepackage{graphicx}	
\usepackage{dcolumn}	
\usepackage{bm}		
\usepackage{amsmath}    
\usepackage{amsfonts}
\usepackage{amssymb}
\usepackage{hyperref}
\usepackage{mathtools}
\usepackage{setspace}	
\usepackage{color}
\usepackage{dsfont}
\usepackage{pdfpages}

\definecolor{green}{rgb}{0,.45,0}
\definecolor{orange}{rgb}{1,0.5,0}

\newcommand{\be}{\begin{equation}}
\newcommand{\ee}{\end{equation}}
\newcommand{\ba}{\begin{eqnarray}}
\newcommand{\ea}{\end{eqnarray}}
\newcommand{\la}{\langle}
\newcommand{\ra}{\rangle}
\newcommand{\di}{\mathrm{d}}

\begin{document}
\title{\boldmath On tetraquarks with hidden charm and strangeness 
as $\phi$-$\psi(2S)$ hadrocharmonium}

\author{Julia Yu.~Panteleeva}
	\affiliation{Physics Department, Irkutsk State University, 
		Karl Marx str.~1, 664003, Irkutsk, Russia}
\author{Irina A.~Perevalova}
	\affiliation{Physics Department, Irkutsk State University, 
		Karl Marx str.~1, 664003, Irkutsk, Russia}
\author{Maxim V.~Polyakov}
	\affiliation{Petersburg Nuclear Physics Institute, 
		Gatchina, 188300, St.~Petersburg, Russia}
	\affiliation{Institut f\"ur Theoretische Physik II, 
		Ruhr-Universit\"at Bochum, D-44780 Bochum, Germany}
\author{Peter~Schweitzer}
	\affiliation{Department of Physics, University of Connecticut, 
		Storrs, CT 06269, USA}
	\affiliation{Institute for Theoretical Physics, T\"ubingen University, 
	Auf der Morgenstelle 14, 72076 T\"ubingen, Germany}

\date{January 2019}  


\begin{abstract} 
In the hadrocharmonium picture a $\bar cc$ state and a light hadron 
form a bound state. The effective interaction is described in terms of 
the chromoelectric polarizability of the $\bar cc$ state and 
energy-momentum-tensor densities of the light hadron. This picture is justified 
in the heavy quark limit, and may successfully account for a hidden-charm 
pentaquark state recently observed by LHCb.
In this work we extend the formalism to the description of hidden-charm
tetraquarks, and address the question of whether the resonant states observed 
by LHCb in the $J/\psi$-$\phi$ spectrum can be described as hadrocharmonia. 
This is a non-trivial question because nothing is known about the 
$\phi$ meson energy-momentum-tensor densities. With rather general 
assumptions about energy-momentum-tensor densities in the $\phi$-meson we show 
that a $\psi(2S)$-$\phi$ bound state can exist, and obtain a characteristic
relation between its mass and width. We show that the tetraquark $X(4274)$
observed by LHCb in $J/\psi$-$\phi$ spectrum is a good candidate for a
hadrocharmonium. We make predictions which will allow testing this picture. 
Our method can be generalized to identify other potential hadrocharmonia.
\end{abstract}

\maketitle

\section{Introduction}
\label{Sec-1:Introduction}

Many evidences for tetraquark states with hidden charm were recently
found, see Refs.~\cite{Ali:2017jda,Olsen:2017bmm,Lebed:2016hpi} for reviews.
In particular, states with hidden strangeness and charm were discovered. 
The most comprehensive analysis of the $J/\psi$-$\phi$ system was performed 
by the LHCb collaboration \cite{Aaij:2016nsc}.  Four tetraquark states 
with quantum numbers $J^{PC}=0^{++},1^{++}$ were observed.

Various theoretical approaches have been suggested to interpret such 
tetraquark states, for instance in terms of hadronic molecules formed 
of $D$-mesons or their excited states 
\cite{Karliner:2016ith,Ding:2009vd,Branz:2009yt} or in the diquark picture 
\cite{Drenska:2009cd,Anisovich:2015caa,Maiani:2016wlq,Anwar:2018sol}.
It was also suggested that the observed structure at $m=4140$~MeV 
is a manifestation of rescattering \cite{Swanson:2015bsa,Liu:2009iw}. 
The fit of the LHCb data on $X(4140)$ in terms of rescattering effects 
in the model of Ref.~\cite{Swanson:2015bsa} gives a slight preference to 
this model over a Breit-Wigner resonance.
The state $X(4274)$ with $J^{PC}=1^{++}$ cannot be described as a molecular 
state or rescattering effect. In \cite{Lu:2016cwr} it was proposed that
$X(4274)$ may be a conventional $\chi_{c1}(3P)$ state. But the couplings of 
$\chi$ charmonia to $J/\psi$-$\phi$ and $J/\psi$-$\omega$ systems
can be naturally 
expected to be similar, and the mass spectrum of $J/\psi$-$\omega$ in 
the decays of $B \to J/\psi\,\omega\,K$ shows no structures analog to those 
in the $J/\psi$-$\phi$ spectrum. This is a strong argument against an
interpretation for any of the states $X(4140)$, $X(4274)$, $X(4500)$,
$X(4700)$ as conventional charmonia. For detailed discussions see 
the reviews \cite{Ali:2017jda,Olsen:2017bmm,Lebed:2016hpi}.

Here we investigate the possibility of whether some of these 
tetraquarks can be interpreted
as bound states of a $\phi$-meson and $\psi(2S)$ in the formalism 
of Refs.~\cite{Voloshin:1979uv,Voloshin:2007dx,Dubynskiy:2008mq}.

This formalism provides a successful description of the pentaquark state 
$P_c(4450)$ observed at LHCb \cite{Aaij:2015tga,Aaij:2016phn,Aaij:2016ymb} 
as a bound state of the nucleon and $\psi(2S)$
\cite{Eides:2015dtr,Perevalova:2016dln} if the chromoelectric 
polarizability of $\psi(2S)$ is $\alpha(2S)\approx 17\,{\rm GeV}^{-3}$.
Lattice data on the $J/\psi$-nucleon potential \cite{Sugiura:2017vks}
support this interpretation \cite{Polyakov:2018aey}.
The formalism makes also predictions for bound states of $\psi(2S)$ 
with $\Delta$ and hyperons \cite{Perevalova:2016dln,Eides:2017xnt} 
which will allow testing this appealing approach in experiment. 
For studies of the $J/\psi$ interaction with nuclear matter 
we refer to \cite{Brodsky:1989jd,Luke:1992tm}.

In this work we investigate whether the hadrocharmonium
picture can also describe some of the hidden-charm tetraquarks.
We will show that the tetraquark $X(4274)$ is a good candidate 
for a bound state of $\psi(2S)$ with a $\phi$-meson. We will also 
make predictions which will allow to test this picture.

\section{The effective quarkonium-hadron interaction}

In the heavy quark limit, when the quarkonium size is much smaller than 
the size of the considered hadron, here $\phi$, the effective interaction 
$V_{\rm eff}$ of an $s$-wave quarkonium with the $\phi$-meson is described 
in terms of the quarkonium polarizability $\alpha$ and the energy-momentum 
tensor (EMT) densities of the $\phi$-meson, 

\vspace{2mm}

\be\label{Eq:Veff}
	V_{\rm eff}(r) = -\,\alpha\;\frac{4\pi^2}{b}\,
	\frac{g_c^2}{g_s^2}\,\biggl(\nu\,T_{00}(r)-3\,p(r)\biggr) 
	, \;\; 
	\nu = 1+\xi_s\,\frac{b\,g_s^2}{8\pi^2}. \;\;
\ee

\vspace{2mm}

\noindent
Here $T_{00}(r)$ and $p(r)$ are the energy density and pressure 
\cite{Polyakov:2002yz} inside the $\phi$-meson, 
which satisfy respectively (see \cite{Polyakov:2018guq} for 
a review on EMT form factors of hadrons and their densities)
\be\label{Eq:mass-von-Laue}
	\int\di^3r\,T_{00}(r)=m_\phi 	\, , \quad \quad
	{\int \di^3 r\,p(r) = 0} 	\, ,
\ee
and $b = (\frac{11}{3} N_c-\frac{2}{3}\,N_f)$ is the leading coefficient 
of the Gell-Mann-Low function, $g_c$ ($g_s$) is the strong coupling 
constant renormalized at the scale $\mu_c$ ($\mu_s$) associated with 
the heavy quarkonium ($\phi$-meson). The parameter $\xi_s$ denotes 
the fraction of the hadron energy carried by gluons at the scale 
$\mu_s$ \cite{Novikov:1980fa}. 
{It is approximately $g_c \approx g_s$ and $\nu\approx 1.5$
\cite{Eides:2015dtr}.}
The derivation of Eq.~(\ref{Eq:Veff}) is justified in the limit that
the ratio of the quarkonium size is small compared to the effective 
gluon wavelength \cite{Voloshin:2007dx}, and a numerically small term 
proportional to the current masses of the light quarks is neglected.

With the value of $\alpha(2S)$ obtained in 
\cite{Eides:2015dtr,Perevalova:2016dln} and a model for EMT
densities, energy density $T_{00}(r)$ and pressure $p(r)$, 
in the $\phi$-meson one in principle is in the position to apply 
the formalism to the description of bound states of $\phi$-mesons 
with $\psi(2S)$. 

It should be remarked that in our situation mixing effects between 
$\bar ss$ and $\bar cc$ components are negligible, because the binding energy 
of a hadrocharmonium is small. In fact, in the heavy quark limit $m_Q\to\infty$ 
the mass of system is of ${\cal O}(m_Q)$ but its binding energy is of 
${\cal O}(m_Q^0)$ and hence much smaller. Thus, in the heavy quark limit, 
which justifies the validity of Eq.~(\ref{Eq:Veff}), mixing effects between 
the light- and heavy-quarkonium components can be consistently neglected.

\section{\boldmath EMT densities in the $\phi$-meson}
\label{Sec:EMTdensities}

Very little is known about the  EMT densities in the $\phi$-meson 
\cite{Abidin:2008ku}. These densities are defined in terms of
Fourier transforms of the EMT form factors $A(t)$ and $D(t)$
\cite{Polyakov:2002yz}. 
The energy density $T_{00}(r)$ and the pressure $p(r)$ entering the 
effective potential (\ref{Eq:Veff}) are expressed in terms of 
form-factors $A(t)$ and $D(t)$ as follows:
\ba
	T_{00}(r) &=& m_\phi \int {\frac{d^3p}{(2\pi)^3}}\ 
	e^{i \bm{p r}}\ A(-\bm{p}^2), \nonumber\\ 
	p(r) &=& \frac{1}{6\,m_\phi}\, \frac{1}{r^2}\frac{d}{dr} r^2\frac{d}{dr} 
	\int {\frac{d^3p}{(2\pi)^3}}\ e^{i \bm{p r}}\ D(-\bm{p}^2).
\ea
Obviously the normalisation conditions (\ref{Eq:mass-von-Laue}) are 
satisfied automatically. We recall that the form factor $A(t)$ 
satisfies the constraint $A(0)=1$, while the value of the $D$-term 
$D=D(0)$ is not fixed \cite{Polyakov:2018guq}.
Almost nothing is known about the $D$-terms of any meson
\cite{Polyakov:2018guq}, except for the recent first phenomenological 
information on $\pi^0$ EMT form factors \cite{Kumano:2017lhr}. But 
$\pi^0$ is a Goldstone boson, and its $D$-term (see \cite{Hudson:2017xug} 
and references therein) does not need to be good guideline for a 
vector meson like $\phi$.

In a very simple description one may {assume simple generic} forms, 
e.g.\ dipole and quadrupole\footnote{We chose the quadrupole Ansatz 
for $D(t)$ in order to avoid a divergent pressure at the origin. 
However, we checked that our results are {only} moderately affected 
if one uses a singular at the origin pressure $p(r)$.} Ans\"atze.
In this case we describe the EMT densities in the $\phi$-meson
in terms of 3 parameters:
\be
\label{Eq:dipolquadrupole}
	A(t) = \frac{1}{(1-t/M_1^2)^2}, \quad \quad
	D(t) = \frac{D}{(1-t/M_2^2)^3}, \quad 
\ee
where $M_1$ is the dipole mass of $A(t)$, $D$ is the value of 
the $D$-term, and  $M_2$ is the quadrupole mass of $D(t)$.
The mass parameter $M_1$ can be related to the mean square radius 
of the energy density in the $\phi$-meson as $r^2_E=12/M_1^2$, 
whereas the mass parameter $M_2$ is related to the mechanical 
mean square radius of the $\phi$-meson (for the definition and 
discussion of the mechanical radius see Ref.~\cite{Polyakov:2018guq})
as $r^2_{\rm mech}=12/M_2^2.$

The radii and $D$-term of the $\phi$-meson are not known 
(see e.g. \cite{Abidin:2008ku}). Therefore here we shall 
assume wide ranges of values for these parameters
(with $i=E,\;{\rm mech}$): 
\be\label{eq:ranges}
 	0.05\  {\rm fm}^2 <r^2_i<1\ {\rm fm}^2, \quad
	-15<D<0.  
\ee
The $D$-term is expected to be negative, see e.g.\ the discussion in 
\cite{Polyakov:2018guq}. The interval of $D$ in (\ref{eq:ranges}) 
includes the value of $D=-1$ which corresponds to the $D$-term for a 
non-interacting point-like vector particle \cite{Holstein:2006ge}.
{In the parameter space (\ref{eq:ranges}) we include on purpose
realistic as well as rather exotic values.}

\begin{figure}[h!]
\centering
\includegraphics[width=6cm]{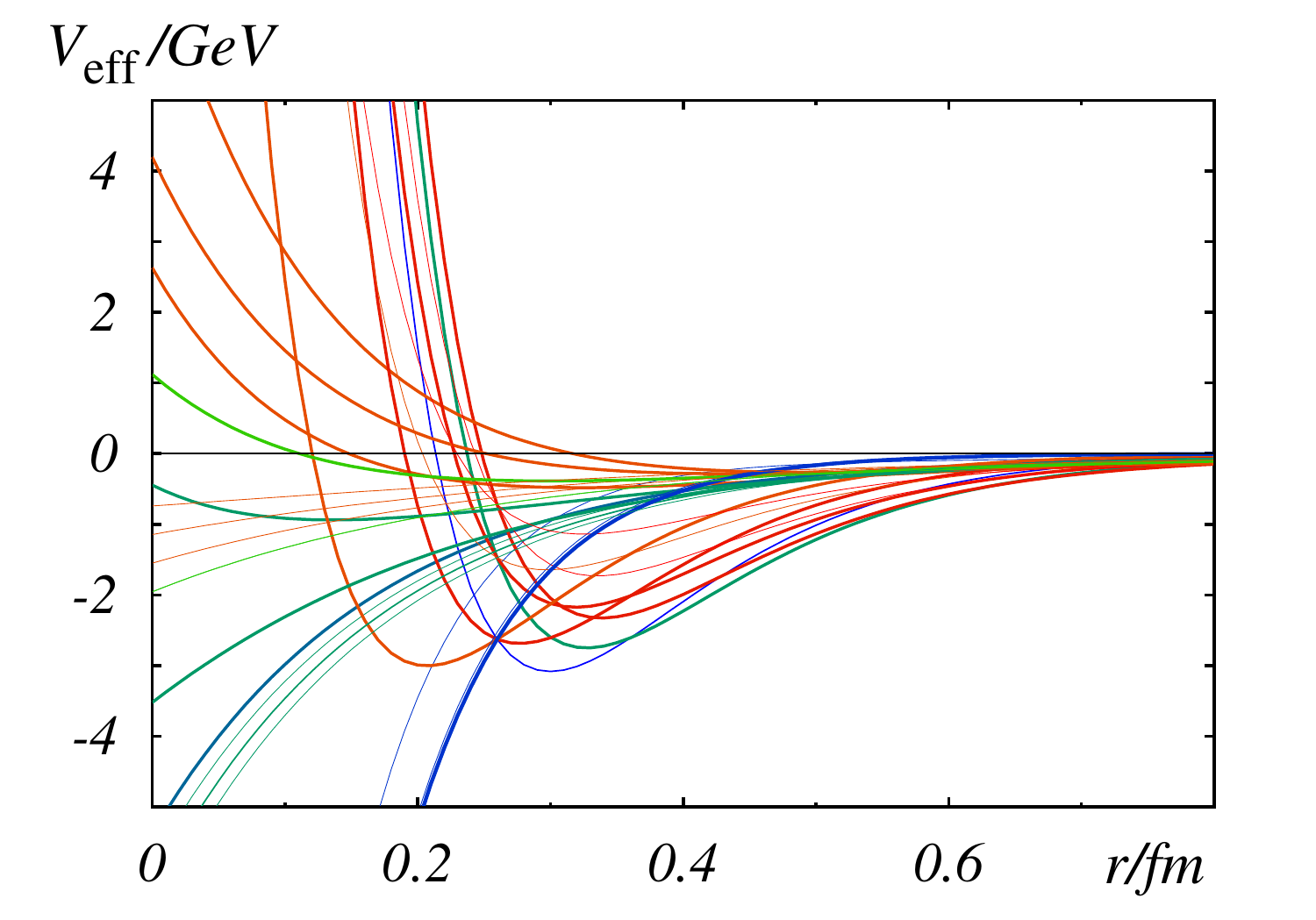}
\caption{\label{Fig:setofVeff}
	Examples of the effective potentials obtained from different 
	values of the parameters in the intervals (\ref{eq:ranges})
	in our Ans\"atze for $\phi$-meson EMT densities.}
\end{figure}

With the parameters in above mentioned intervals we obtain a set of effective 
potentials whose form varies considerably. For illustrative purposes we plot 
in Fig.~\ref{Fig:setofVeff} examples of the resulting effective potentials.
Due to the normalisation conditions (\ref{Eq:mass-von-Laue}) all 
effective potentials in the set are normalised by the condition:
\be\label{Eq:Veff-norm}
	\int\di^3r\;V_{\rm eff}(r) = -\,\alpha\;
	\frac{4\pi^2}{b}\,\frac{g_c^2}{g_s^2}\;\nu\,m_\phi. 
	\ee
In the next sections we study the possible $\psi(2S)$-$\phi$ bound states 
and their partial decay width to $\phi$ and $J/\psi$.

\section{\boldmath Mass and partial decay width of the $\psi(2S)$-$\phi$ 
hadrocharmonium }

Let $m_\psi$, $m_J$, $m_\phi$ denote the masses of $\psi(2S)$, 
$J/\psi$, $\phi$-meson. The mass of the tetraquark state is defined 
as $M=m_\psi+m_\phi+E_{\rm bind}$. The binding energy $E_{\rm bind} < 0$ is 
obtained from solving the non-relativistic Schr\"odinger equation 
with the effective potential defined in terms of the $\psi(2S)$ 
chromoelectric polarizability $\alpha(2S)$ \cite{Eides:2015dtr}
\be\label{Eq:Schroedinger}
	\left(-\frac{\bm{\nabla}^2}{2\mu_2}+V_{\rm eff}(r)-E_{{\rm bind}}
	\right)\Psi(\bm{r})=0,
\ee
where $\mu_2$ is the reduced mass $\mu_2^{-1} = m^{-1}_\psi+m^{-1}_{\phi}$
of the bound particles.

The decay of the tetraquark into $\phi$ and $J/\psi$ {requires that 
$M>m_J+m_\phi$ and} is governed by the same effective potential but 
rescaled, since now the $\alpha(2S\to1S)$ polarizability is relevant.
The formula for the decay width is given by \cite{Eides:2015dtr,Eides:2017xnt}
\be\label{eq:width}
	\Gamma = \frac{\mu_1|\bm{q}|}{\pi}\;
	\biggl(\frac{\alpha(2S\to1S)}{\alpha(2S)}\biggr)^{\!\!2}\;
	\Biggl|\int\di^3r\;\Psi(\bm{r})\,V_{\rm eff}(r)\,e^{i\bm{qr}}\Biggr|^2
\ee
where $\mu_1$ is the reduced mass $\mu_1^{-1} = m^{-1}_J+m^{-1}_{\phi}$
of the decay products, and $|\bm{q}| = \sqrt{2\mu_1 (M-m_J-m_\phi)}$ 
corresponds to the center-of-mass frame momentum of the decay products. 
The bound-state wave function $\Psi(\bm{r})$ corresponding to the binding 
energy $E_{\rm bind} =M-m_\psi-m_\phi$ is normalised to unity, 
$\int d^3r\  |\Psi(\bm{r})|^2=1$.

To evaluate the binding energy and width in 
Eqs.~(\ref{Eq:Schroedinger},~\ref{eq:width}) we use the value 
$\alpha(2S)\approx 17\,{\rm GeV}^{-3}$ which was shown to yield a 
robust description of the pentaquark state $P_c(4450)$ interpreted as a
$N$-$\psi(2S)$ bound state under varying assumptions of different chiral 
models for nucleon EMT densities \cite{Eides:2015dtr,Perevalova:2016dln}. 
  In a recent study \cite{Anwar:2018bpu} a wide range of values was 
  estimated $18\,{\rm GeV}^{-3}\lesssim\alpha(2S)\lesssim270\,{\rm GeV}^{-3}$
  by inferring $\alpha(1S)$ from available results for 
  nucleon-$J/\psi$ scattering~lengths and exploring the 
  relation $\alpha(2S)/\alpha(1S)=502/7$ derived in the heavy-quark 
  and large-$N_c$ limit by treating quarkonia as Coulomb systems 
  \cite{Peskin:1979va}. Interestingly, the lowest value of this 
  range is compatible with $\alpha(2S)\approx 17\,{\rm GeV}^{-3}$ 
  from \cite{Eides:2015dtr,Perevalova:2016dln}.
For the transitional chromoelectric polarizability we use 
$|\alpha(2S\to1S)|\approx 2$GeV$^{-3}$ from Ref.~\cite{Voloshin:2007dx}.
The $\phi$-meson EMT densities are modeled as described in 
Sec.~\ref{Sec:EMTdensities} with parameters varied in the 
wide intervals of Eq.~(\ref{eq:ranges}). 

Not surprisingly, we obtain a wide range of masses $M$ for the 
corresponding tetraquarks: practically every $M$ in the allowed 
range $m_J+m_\phi < M < m_\psi+m_\phi$ is realized for some choices
of parameters $M_1$, $M_2$, $D$ in the range (\ref{eq:ranges}). 
Also the results for $\Gamma$ vary considerably. 

\begin{figure}[b!]
\centering
\includegraphics[width=7cm]{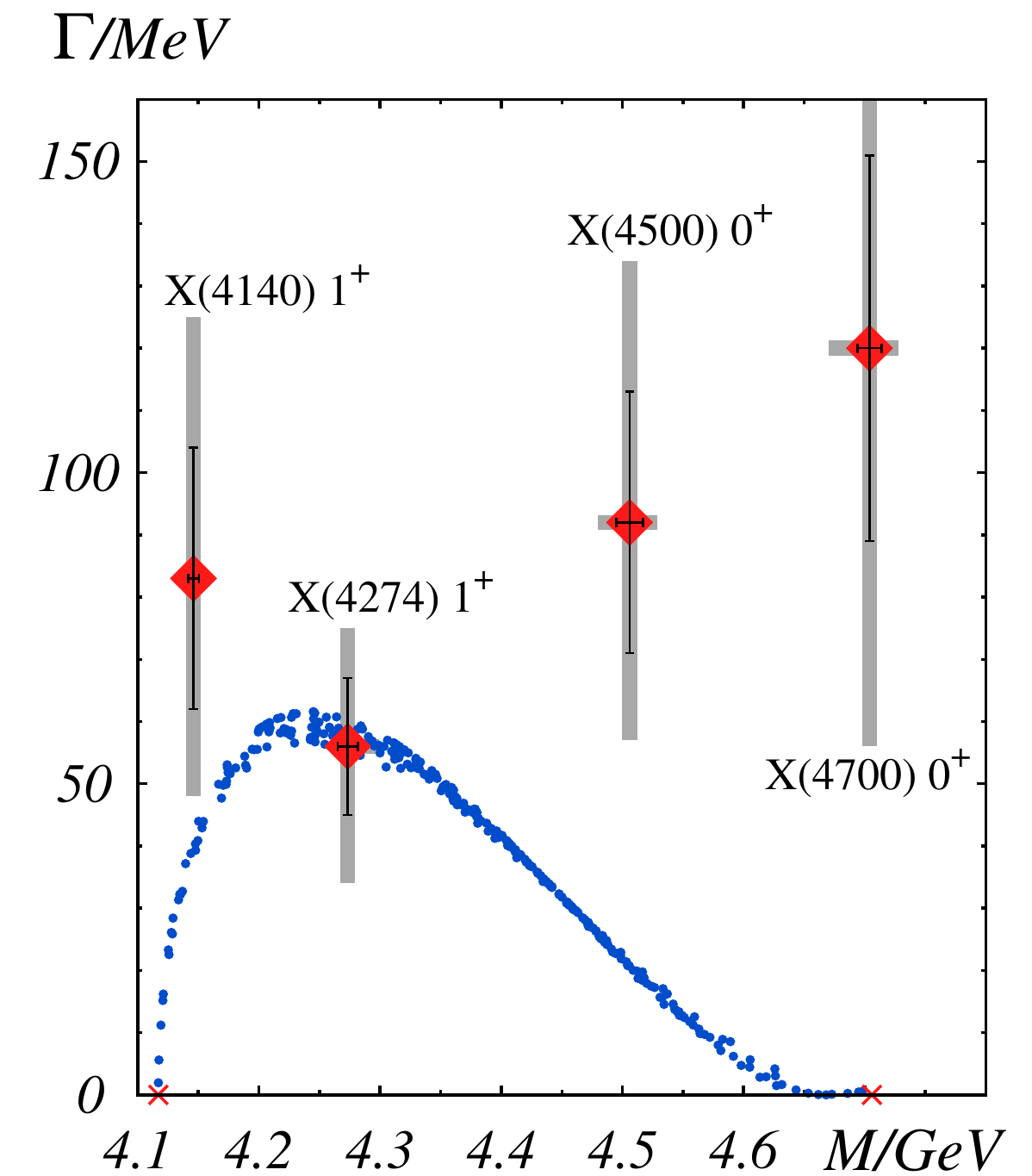}
\caption{\label{FIG-01:observation-picture}
	The scatter plot of the decay width $\Gamma(M_1,M_2,D)$ vs 
	mass $M(M_1,M_2,D)$ of tetraquarks obtained from varying the
	parameters $M_1,\,M_2,\,D$, which describe the unknown $\phi$-meson 
	EMT form factors (\ref{Eq:dipolquadrupole}), within a wide range 
	of the values {(\ref{eq:ranges}). In this plot 310 different
	points are shown! Remarkably, even though we randomly scan a 
	large parameter space, the $\Gamma$-$M$-values lie approximately 
	on a characteristic curve, see text. The crosses on 
	the $M$-axis indicate the bounds $m_J+m_\phi < M < m_\psi+m_\phi$.
	For comparison we show the four tetraquarks in the $J/\psi$-$\phi$ 
	resonance region with their statistical (thin lines) and systematic 
	(shaded areas) uncertainties and spin parity assignments 
	\cite{Aaij:2016nsc}. The state $X(4274)$ emerges as a candidate for 
	the description as a hadrocharmonium.
	This method can be used to identify other possible hadroquarkonia.}}
\end{figure}

The mass and width are functions $\Gamma(M_1,M_2,D)$ and $M(M_1,M_2,D)$
of parameters $M_1$, $M_2$, $D$ which are varied randomly in the ranges 
(\ref{eq:ranges}). At first glance one would expect a scatter plot 
of $\Gamma(M_1,M_2,D)$  versus $M(M_1,M_2,D)$ to yield a random 
$\Gamma$-$M$-distribution filling out the whole $M$--$\Gamma$ plane. 
But surprisingly we find that the points lie more or less on one curve, 
see Fig.~\ref{FIG-01:observation-picture}. This is remarkable: even 
though we know nothing about the structure of the $\phi$-meson, we can 
predict that $M$ and $\Gamma$ of candidate $\psi(2S)$-$\phi$ tetraquarks 
are systematically correlated.
This is not a feature of a particular parametrization (dipole and quadrupole). 
We checked that it is also the case for other form factor parametrizations, 
e.g.\ higher multipoles. Very similar results are obtained also
with  EMT densities of a ``smeared out'' point-like boson 
\cite{Hudson:2017xug} or with a simple square well potential. 
Notice that the same values for ($\Gamma$,~$M$) can be obtained from 
different combinations of the parameters in the intervals (\ref{eq:ranges}).

In the remainder of this section we will clarify the question why $M$ 
and $\Gamma$ are correlated in this characteristic way. In the next section
we will address the implications of this finding.

The bound state problem and the width can be conveniently solved and
evaluated in position space. To understand the $\Gamma$-$M$-relation
it is convenient to work in momentum space. Assuming that the bound 
state problem is solved (in position space) and the wave function 
$\Psi(\bm{r})$ is known, we define the momentum-space wave function as
\be
	\widetilde\Psi(\bm{p})= \int\di^3r\;e^{-i\bm p r}\,\Psi(\bm{r})\,,
\ee
and introduce the form factor $F_{\rm eff}(\bm{p})\equiv F_{\rm eff}(-\bm{p}^2)$
as the Fourier transform of the effective potential as
\be
	V_{\rm eff}(r) =  \int\frac{\di^3p}{(2\pi)^3}\;F_{\rm eff}(\bm{p})\
	e^{i\bm{p r}} \,.
\ee
If we take the Schr\"odinger equation in momentum space
\be\label{Eq:Schroedinger-mom}
	\left(\frac{\bm{p}^2}{2\mu_2}-E_{{\rm bind}}\right)\widetilde\Psi(\bm{p})
	= - \int\frac{\di^3p^\prime}{(2\pi)^3}\;F_{\rm eff}(\bm{p}-\bm{p^\prime})
	\widetilde\Psi(\bm{p^\prime})
\ee
and multiply it by its complex conjugate, we obtain
\be\label{Eq:Schroedinger-mom-square}
	\left(\frac{\bm{p}^2}{2\mu_2}-E_{{\rm bind}}\right)^{\!\!2}
	|\widetilde\Psi(\bm{p})|^2
	= 
	\Biggl|\int\frac{\di^3p^\prime}{(2\pi)^3}\;F_{\rm eff}(\bm{p}-\bm{p^\prime})
	\widetilde\Psi(\bm{p^\prime})\Biggr|^2\,.
\ee
At the same time, the formula for the decay width can be expressed as
\be
	\Gamma = 
	\biggl(\frac{\alpha(2S\!\to\!1S)}{\alpha(2S)}\biggr)^{\!\!2}\
	\frac{\mu_1|\bm{q}|}{\pi}
	\Biggl|\int\frac{\di^3p^\prime}{(2\pi)^3}\,F_{\rm eff}(\bm{q}-\bm{p^\prime})
	\widetilde\Psi(\bm{p^\prime})\Biggr|^2\!.\;
\ee
Thus we see that the binding energy and the partial width of the
hadrocharmonium are related as
\be\label{Eq:width-vs-mass}
	\Gamma =
	\biggl(\frac{\alpha(2S\!\to\!1S)}{\alpha(2S)}\biggr)^{\!\!2}\;
	\frac{\mu_1|\bm{q}|}{\pi}
	\left(\frac{\bm{q}^2}{2\mu_2}-E_{{\rm bind}}\right)^{\!\!2}
	|\widetilde\Psi(\bm{q})|^2, 
\ee
with $|\bm{q}| = \sqrt{2\mu_1 (E_{\rm bind}+m_\psi-m_J)}$. Notice
that the center-of-mass momentum of the decay products is 
bound as $0 < \bm{q}^2 < 2\mu_1 (m_\psi-m_J)$.

Consider a class of potentials obtained from 
continuously-differentiable (adiabatic) variations of certain 
parameters. Then $\widetilde\Psi(\bm{p})$, and hence also 
$|\widetilde\Psi(\bm{q})|^2$, will vary in a continuously
differentiable manner as the parameter space is scanned. 
If we varied a single parameter in a potential, we would 
obtain a unique $\Gamma$-$M$-curve.
In our case we vary multiple parameters in the potential, and
obtain families of  $\Gamma$-$M$-curves. Notice, however, that 
only those deformations of $V_{\rm eff}(r)$ are possible which
preserve the normalization condition (\ref{Eq:Veff-norm}). 
This explains why the results for ($\Gamma$,~$M$) all occupy 
a relatively narrow region in the $\Gamma$-$M$ plane. 

The specific shape of the $\Gamma$-$M$-curves can be understood as
follows. For $M\to m_J+m_\phi$ we have $|\bm{q}|\to 0$, i.e.\ the
phase space of the decay naturally suppresses the decay width
as $\Gamma = c_1\,|\bm{q}|$ for small $|\bm{q}|$.
The dimensionless coefficient $c_1$ is of order unity and weakly 
dependent on the details of the wave functions, see App.~\ref{App:A}.
In the opposite limit $M\to m_\psi+m_\phi$ we deal with a bound state 
problem in the threshold limit $E_{\rm bind}\to 0$. In a weakly bound 
case many properties of a quantum system are largely insensitive to 
the details of the specific potential, see e.g.\ the pioneering work 
of Wigner on deuteron \cite{Wigner}. {This implies a suppression 
of the momentum-space wave function in the limit of $E_{\rm bind}\to 0$,
such that $\Gamma$ approaches zero, see Appendix~\ref{App:A}.}

In summary, in the hadrocharmonium picture the mass and partial width 
$\Gamma$ of a tetraquark decaying into $J/\psi$ and $\phi$ are
correlated in a characteristic way.

\section{Discussion of results and conclusions}

The EMT densities in the $\phi$-meson are not known. This prevents us 
from making explicit predictions for the mass of the $\psi(2S)$-$\phi$
bound state in the hadrocharmonium picture. With physically 
very broad assumptions about the EMT densities in the 
$\phi$-meson and taking the value of the chromoelectric polarizability 
of $\psi(2S)$ to be $\alpha(2S)\approx 17\,{\rm GeV}^{-3}$ as needed to 
describe $P_c(4450)$ pentaquark as a bound state of the nucleon and 
$\psi(2S)$ \cite{Eides:2015dtr,Perevalova:2016dln}, we obtained that 
a $\psi(2S)$-$\phi$ bound state can form. Although we cannot make 
precise predictions for the mass of such state, we obtained 
a characteristic relation between mass of the state and 
its partial decay width to $J/\psi$ and $\phi$.

In our approach the $s$-wave bound state of the two vector mesons
$\psi(2S)$ and $\phi$ with $J^{PC}=1^{--}$ has positive parity and 
positive $C$-parity, and corresponds to a mass-degenerate 
multiplet $J^{PC}=0^{++},1^{++}, 2^{++}$. The degeneracy is lifted 
by the hyperfine interaction which is suppressed by the inverse of the
heavy quark mass and expected to be small. Recent lattice studies of the
$J/\psi$-$N$ effective potentials  \cite{Sugiura:2017vks} showed 
that the hyperfine interaction is very small.

Interestingly, the state $X(4274)$ observed in the $J/\psi\;\phi$ channel 
has a width of $\Gamma=56\pm 11^{+8}_{-11}$~MeV \cite{Aaij:2016nsc} exactly 
in the range predicted by our scatter plot, see 
Fig.~\ref{FIG-01:observation-picture}. The LHCb collaboration obtained for 
this state the quantum numbers $J^{PC}=1^{++}$. If one interprets this state
as a $\psi(2S)$-$\phi$ bound state, one should expect two further nearly 
mass-degenerate resonances with spin 0 and 2 in this energy region. 
It would be interesting to check this hypothesis in partial wave analysis.  

It is important to stress that adopting this interpretation for 
$X(4274)$ implies that the 
$X(4140)$, $X(4500)$, $X(4700)$ cannot be $s$-wave $\psi(2S)$-$\phi$ 
bound states. These states could be other hadrocharmonium states, 
possibly with $l\ge1$ which might be possible in specific regions
of the parameter space. Or their explanation may require different 
binding mechanisms. Addressing this question goes beyond the scope
of this work.

Assuming that the state $X(4274)$ is a hadrocharmonium allows us
to gain some (very vague) information on the EMT densities of the
$\phi$-meson.
The  $\psi(2S)$-$\phi$ bound state with the mass around $X(4274)$ appears 
for the following range of parameters $r^2_E \in [0.1, 0.55]~$fm$^2$, 
$r^2_{\rm mech}\in [0.08, 0.5]~$fm$^2$ and $D \in [-5,0]$, the smaller 
radii correspond the larger values of $|D|$. This is a very reasonable 
range of parameters for EMT densities in the $\phi$-meson: for example, 
in the AdS/QCD model one finds $r_E^2=0.21$~fm$^2$ for the
$\rho$-meson \cite{Abidin:2008ku}. This approach would yield  
similar results for other vector mesons such as $\phi$.

We also note that if we consider the chromoelectric polarizability 
$\alpha(2S)$ as a free parameter, the $\psi(2S)$-$\phi$ bound state appears 
for $\alpha(2S)\gtrsim\alpha_{\rm crit}(2S)\in [2,4]$~GeV$^{-3}$ if we 
vary the parameters of EMT densities in above mentioned range. Note that 
this range of critical values for the chromoelectric  polarizability is 
just slightly above the polarizability $\alpha(1S)=1.5\pm 0.6$~GeV$^{-3}$ 
of $J/\psi$ determined in Ref.~\cite{Polyakov:2018aey} from the lattice data 
of Ref.~\cite{Sugiura:2017vks}. 
We remark that $\alpha_{\rm crit} (1S)$ is larger (for the same potential) 
than $\alpha_{\rm crit} (2S)$ due to $\mu_1<\mu_2$. Thus, bound states of 
$J/\psi$ and $\phi$ most probably are not possible in the 
hadrocharmonium picture. This is in line with lattice QCD studies
where the $J/\psi$-$\phi$ potential was found too weak to form bound 
states \cite{Ozaki:2012ce}.

Using the example of the $\psi(2S)$-$\phi$ hadrocharmonium we demonstrated 
that the partial $J/\psi$-$\phi$ decay width is correlated in a 
characteristic way with the mass of the state. This interesting 
``approximate universality'' of the $\Gamma$-$M$ dependence is a generic
feature of the approach and can be expected to hold also for other 
hadroquarkonia. The implications of this observation will be 
studied elsewhere. 

Other interesting questions concern whether 
also other $J/\psi$-$\phi$ resonances can be described as bound or 
resonant states in the hadrocharmonium picture, and whether 
hadroquarkonia with the heavier $\bar bb$ states can exist.
The chromoelectric polarizabilities of bottomonia are
smaller than for charmonia \cite{Brambilla:2015rqa}, and
the corresponding $V_{\rm eff}$ is in general weaker. The
formation of hidden-bottom  tetraquarks in the hadrocharmonium picture
may therefore be more difficult.
But these interesting topics deserve dedicated studies and will be 
addressed elsewhere.

\section{Acknowledgments}

JP, IP and MVP are thankful to Prof. S.~E.~Korenblit for useful discussions. 
The work of MVP is supported by CRC110 (DFG).
This work was supported in part by the National Science Foundation 
(Contract No.~1406298 and 1812423) and the Wilhelm Schuler Stiftung.

\appendix 

\section{\boldmath The partial decay width $\Gamma$ in extreme limits}
\label{App:A}

In the limit $|\bm{q}|\to 0$, where $E_{\rm bind}\to m_J-m_\psi$
approaches its maximal value, we obtain from (\ref{Eq:width-vs-mass}) 
\ba\label{App-eq:1}
	\Gamma &=& c_1\,|\bm{q}| + {\cal O}(|\bm{q}|^3)\, , \nonumber\\
	c_1& =& 
	\biggl(\frac{\alpha(2S\to1S)}{\alpha(2S)}\biggr)^{\!\!2}\;
	4\mu_1(m_\psi-m_J)^2\;\la r^{3/2}\ra^2 \,.
\ea
In Eq.~(\ref{App-eq:1}) we defined 
$\la r^{3/2}\ra = \int_0^\infty\di r\,r\, u(r)$. Here $u(r)$ is the
radial part $u(r)$ of the ($s$-wave) ground-state wave function 
$\Psi(\bm{r}) = u(r)/r\,Y_{00}$. We define $u(r)$ to be real, 
positive, and normalized as $\int_0^\infty\di r \,u(r)^2 = 1$. 
Notice that $u(r)$ has dimension (length)$^{-1/2}$. One has
naturally $\la r^{3/2}\ra^2 = a_0\,R_h^3$. Here $R_h$ is the
characteristic hadronic radius of the problem associated with 
the range of the potential $V_{\rm eff}(r)$ and set by the radius 
of the $\phi$-meson, and $a_0$ is a numerical factor of order unity. 
Quark models  indicate that the $\phi$-meson is about the size 
of the proton or somewhat smaller. If we use this as a guideline 
and assume for the characteristic radius $R_h\sim 0.8\,{\rm fm}$,
we find for the slope $c_1 \sim 1$.

In the opposite limit when $E_{\rm bind}\to 0$ the size of the 
bound-state wave function {in coordinate space} grows as 
$\sim 1/\sqrt{2 \mu_2 |E_{\rm bind}|}$. This implies that the 
momentum-space wave function $\widetilde \Psi(\bm{p})$ becomes 
more and more narrow and hence $|\widetilde\Psi(\bm{q})|^2$ in 
Eq.~(\ref{Eq:width-vs-mass}) goes to zero for fixed $\bm{q}$.
One can show on general grounds (see e.g.\ Ref.~\cite{BZP}) that 
in the limit $E_{\rm bind}\to 0$ and $\bm{q}$ fixed, 
the wave function (squared) in the momentum space 
$|\widetilde\Psi(\bm{q})|^2 \propto \sqrt{-E_{\rm bind}}$ 
and hence $\Gamma \propto \sqrt{-E_{\rm bind}}$. 

\

\end{document}